\documentclass[aps,preprint]{revtex4}%
\usepackage{amsfonts}
\usepackage{amsmath}
\usepackage{amssymb}
\usepackage{graphicx}%
\begin{document}
\title[Hidden Momentum]{Concerning Hidden Momentum}
\author{Timothy H. Boyer}
\affiliation{Department of Physics, City College of the City University of New York, New
York, New York 10031}

\begin{abstract}
The fact that the author of an excellent textbook on electromagnetism could be
duped by "hidden momentum" vividly illustrates the problematic nature of its use.

\end{abstract}
\maketitle

It greatly distresses me that the ideas of "hidden momentum,"\cite{CV}%
\cite{APV}\cite{V} which are of dubious value in the physics research
literature, should now appear in discussions of dubious value or of outright
error within respectable textbooks of classical electromagnetism.\cite{G}%
\cite{J}  It seems that the invocation of "hidden momentum" is simply a
confusing distraction from the needed discussions of energy and momentum flow
within systems.\cite{B}

As an example of the outright errors caused by "hidden momentum," consider the
discussion where "hidden momentum" is first invoked in an excellent
undergraduate electromagnetism text.\cite{ex} In the illustrative example
treated in this text, energy is transferred from a battery on the left to a
resister on the right by way of a coaxial cable. The text calculates the
Poynting vector, the total electromagnetic power flow down the cable, and the
total electromagnetic linear momentum located in the coaxial cable. Clearly
the Poynting vector is associated with the transfer of energy and momentum
from the battery (which loses energy), through the coaxial cable (which does
not change), to the resistor (which gains energy). The battery clearly recoils
on sending out the energy and becomes less massive, while the resistor recoils
and becomes more massive on absorbing the energy. In the presentation of the
textbook, it is not clear how the battery, coaxial cable, and resistor are
connected together. If they are all mounted on a board with wheels, then (in
response to the coupling of the recoils) we would expect the board would roll
to the left so as to maintain the system center of energy at a constant
location in an inertial frame where the system was initially at rest.\cite{T}

However, rather than discussing these aspects of energy flow, the text invokes
the mysteries of "hidden momentum." The text states: "In this case it turns
out that there is 'hidden' mechanical momentum associated with the flow of
current, and this exactly cancels the momentum in the fields." The text also
makes an appeal to an irrelevant later example of "hidden momentum" where no
battery and resistor are present. Thus the text makes the erroneous suggestion
that the \textit{electrical currents} in the cable carry a "hidden mechanical
momentum" in the opposite direction from the electromagnetic energy flow. This
is not the case.

How do we know that that the currents in the coaxial cable do not carry the
mysterious "hidden mechanical momentum" invoked by the text? I believe that we
can make the essential physics of the situation clearer by making the current
carriers less obscure. Let's replace the current-carrying coaxial cable by two
frictionless parallel filaments on which charged beads of negligible mass are
allowed to slide freely. There are positively charged beads lined up along the
top filament and negatively charged beads lined up on the bottom filament,
leading to an electrostatic potential between the filaments. A source of
energy at the left takes a pair of beads (one positive and one negative) and
moves them apart, pushing the positively charged bead onto the top filament
and the negatively charged bead onto the bottom filament. On doing so, the
source of energy both does work and also recoils to the left. The energy is
absorbed at the right-hand end of the filaments where a positive bead is
removed from the top filament while a negative bead is removed from the bottom
filament, and energy is transferred as the beads are removed and brought
together. The energy and momentum flow down the pair of filaments is purely
electromagnetic and is given by Poynting's vector, and there is
electromagnetic momentum stored in the region of the filaments. However, there
is nothing even remotely resembling "hidden mechanical momentum" in the
currents on the filaments. This example suggests that the comments in the text
claiming the existence of "hidden mechanical momentum" associated with the
flow of current are simply wrong.

Perhaps the situation is even more transparent if we consider electromagnetic
energy transmitted through vacuum from the energy source on the left to an
energy receiver on the right in the form of electromagnetic waves. Once again
the energy source on the left will lose energy while the energy sink on the
right will gain energy, and the space in between source and sink will contain
both electromagnetic energy and momentum. In this case there are no electrical
currents for which one can invoke "hidden mechanical momentum." Of course, in
all these cases, if the energy source and energy sink are mounted on a board
on wheels, the board will roll to the left due to the recoils associated with
energy transmission and reception.\cite{T}

The fact that the author of an excellent textbook on electromagnetism could be
duped by "hidden momentum" vividly illustrates the problematic nature of its use.

\end{document}